\documentclass[prb,showpacs,amsmath,amssymb,twocolumn,a4paper]{revtex4}
\usepackage[latin1]{inputenc}

\usepackage{graphicx,amsmath,amssymb, amsfonts}
\begin{document}
\title{Kinetic Vlasov Simulations of collisionless magnetic Reconnection}
       
\author{H. Schmitz and R. Grauer}

\affiliation{
Theoretische Physik I, 
Ruhr-Universit\"at, 44780 Bochum, Germany }


\begin{abstract}

A fully kinetic Vlasov simulation of the Geospace Environment Modeling
(GEM) Magnetic Reconnection Challenge is presented. Good agreement is found
with previous kinetic simulations using particle in cell (PIC) codes,
confirming both the PIC and the Vlasov code. In the latter the complete
distribution functions $f_k$ ($k=i,e$) are discretised on a numerical grid
in phase space. In contrast to PIC simulations, the Vlasov code does not
suffer from numerical noise and allows a more detailed investigation of the
distribution functions. The role of the different contributions of Ohm's
law are compared by calculating each of the terms from the moments of the 
$f_k$. The important role of the off--diagonal elements of the electron
pressure tensor could be confirmed. The inductive electric field at the
X--Line is found to be dominated by the non--gyrotropic electron pressure,
while the bulk electron inertia is of minor importance. Detailed analysis
of the electron distribution function within the diffusion region reveals
the kinetic origin of the non--gyrotropic terms.

\end{abstract}




\maketitle

\section{Introduction}

Magnetic reconnection  is the fundamental process which allows magnetized
plasmas to convert the energy stored in the field lines into kinetic energy
of the plasma. It plays an important role in the dynamics of space and
laboratory plasmas. In the magnetopause it allows particles from the solar
wind to enter the magnetosphere. Also it is believed to be the main source
of energy for solar flares and coronal mass ejections.

In ideal magnetohydrodynamics (MHD) the frozen--in flux condition prohibits
the magnetic field topology to change. Thus reconnection depends on a
mechanism that breaks the frozen--in condition. This non--ideal mechanism
is responsible for the dynamics of the diffusion region, where the topology
change takes place. In resistive MHD with low values of resistivity an
elongated Sweet--Parker current sheet develops, which limits the
reconnection rate. \cite{Parker:57,Sweet:58,Biskamp:86} Higher values for
the resistivity, on the other hand, are extremely unrealistic for the
astrophysical reconnection processes. 

In the last years it has become apparent that the minimal model to
adequately describe collisionless reconnection is Hall--MHD.
\cite{Shay:01,Lottermoser-Scholer:1997,Wang-etal:2001,Huba-Rudakov:2004} In
Hall--MHD the ion inertial length $\lambda_i$ is introduced as a
characteristic scale length such that for distances smaller than
$\lambda_i$ the ion dynamics can decouple from the magnetic field. Within
the framework of the Geospace Environmental Modeling (GEM) Magnetic
Reconnection Challenge a two dimensional magnetic reconnection setup based
on the Harris equilibrium was studied. Simulations were done using MHD
models with constant or localized resistivity \cite{Birn:2001b,Otto:2001},
Hall--MHD models \cite{Birn:2001b,Otto:2001,Ma:2001}, hybrid models where
electrons are treated as fluid and ions are treated kinetically
\cite{Shay:01,Kuznetsova:2001} and with fully kinetic particle codes.
\cite{Shay:01,Pritchett:2001b,Hesse:2001} The models which included the
Hall effect, either directly or implicitly in the kinetic treatment, showed
very similar results in terms of the global reconnection rate. The small
scale structures of the dissipation region, on the other hand, varied
substantially between the models. In Ref.\ \onlinecite{Birn:2001a} it was
concluded  that the large scale ion dynamics control the rate of
reconnection and that the dispersive character of Whistler modes is
essential in the understanding of the results.

These results emphasize the importance of the Hall term for the
reconnection. At the neutral X--line, the magnetic field strength is
exactly zero and the Hall term vanishes. Only the electron bulk inertia and
the non--gyrotropic part of the electron pressure tensor can yield a
contribution at the X--line. Kuznetsova {\it et al.} \cite{Kuznetsova:1998} pointed
out, that the bulk inertia of the electrons can only contribute if the
current sheet develops on electron timescales. Vasyliunas
\cite{Vasyliunas:1975} was the first to emphasize the important role of the
off--diagonal components of the pressure tensor in the diffusion region.
Later their role was investigated by a number of authors (see e.g
Refs.\ \onlinecite{Dungey:1988,Hesse:1993,Kuznetsova:1998,Kuznetsova:2001}). The
non--gyrotropic pressure can be understood as originating from the
meandering motion of the electrons, which bounce in the magnetic field
reversal region, also known as Speiser motion. \cite{Speiser:1991}

In this study, we use a Vlasov code to investigate the GEM reconnection. In
contrast to Particle--In--Cell (PIC) codes, Vlasov codes discretise the
complete distribution functions of electrons and ions on a grid in phase
space. This has the advantage of allowing more detailed investigations of
the distribution functions, since numerical noise is completely absent from
the Vlasov--approach. The main focus of our investigation is the electron
dynamics. These can be understood in two ways. On one hand, one can take
a microscopic point of view by looking at the detailed structure of
the distribution function. On the other hand one can look at the
moments of the distribution function and their importance for Ohm's law.
The Vlasov--approach allows us to do both with very high accuracy.

In the next section we briefly present the methods used in our
investigation. The GEM reconnection setup including the initial conditions
and the boundary conditions is described in section \ref{SecSetup}. In
section \ref{SecResults} we discuss the results. Separate subsections are
dedicated to the discussion of the contributions of the terms in Ohm's law,
especially the off--diagonal components of the pressure tensor, and to the
detailed discussion of the electron distribution function. Section
\ref{SecSummary} will give a summary and present conclusions.

\section{Methods\label{SecMethods}}

The basis of the kinetic description of magnetic reconnection are the
distribution functions $f_k({\bf x}, {\bf v},t)$ of species $k$, where
$k=i,e$ denotes ions or electrons. The temporal behavior of the
distribution function is given by the Vlasov equation
\begin{displaymath}
\frac{\partial f_k}{\partial t} 
        + {\bf v}\cdot \nabla f_k
        + \frac{q_k}{m_k} \left( 
                {\bf E} + {\bf v} \times {\bf B}
        \right) \cdot \nabla_{\bf v} f_k 
= 0.\label{VlasovOrig}
\end{displaymath}
Here $q_k$ and $m_k$ are the charge and the mass of the particles of species
$k$. The Vlasov equation describes the incompressible flow of the species
phase space densities under the influence of the electromagnetic fields.

The fields are solved using the Darwin approximation (see, for example
Refs.\ \onlinecite{BIR85,SCHM06a})
\begin{alignat}{1}
  \nabla\times\mathbf{E}_T &= -\frac{\partial \mathbf{B}}{\partial t} \;\; ,\\
  \frac{1}{\mu_0}\nabla\times\mathbf{B} &=
  \varepsilon_0 \frac{\partial \mathbf{E}_L}{\partial t} + {\bf j} \;\; ,\\
  \nabla\cdot\mathbf{E}_L &= \frac{1}{\varepsilon_0} \rho \;\; ,\\
  \nabla\cdot\mathbf{B} &= 0 \;\; ,
\end{alignat}
where $\mathbf{B}$ is the magnetic and $\mathbf{E}_L$ and $\mathbf{E}_T$
are the longitudinal and the transverse part of the electric field,
$\nabla\times\mathbf{E}_L=0$, $\nabla\cdot\mathbf{E}_T=0$, $\mathbf{E} =
\mathbf{E}_L + \mathbf{E}_T$. The Darwin approximation eliminates the fast
electromagnetic vacuum modes while all other plasma modes are still
described. Instead of neglecting the displacement current completely, only
its transverse part is dropped while its longitudinal part is kept. The
elimination of the vacuum modes allows larger time-steps in the simulation
since only the slower non--relativistic waves have to be resolved.

The charge density $\rho$ and the current density $\mathbf{j}$ are given by
the moments of the distribution function,
\begin{alignat}{1}
\rho &= \sum_k q_k \int f_k(\mathbf{x}, \mathbf{v}) d^3v \;\; ,\\
\mathbf{j} &= \sum_k q_k \int \mathbf{v} f_k(\mathbf{x}, \mathbf{v}) d^3v \;\; .
\end{alignat}
These moments couple the electromagnetic fields back to the distribution
function. In this way the Vlasov-Darwin system constitutes a non--linear
system of partial integro--differential equations.

For the simulations we use a $2\frac{1}{2}$--dimensional Vlasov--code
described in Ref.\ \onlinecite{SCHM06a}. Here $2\frac{1}{2}$--dimensional 
means, we restrict the simulations to 2 dimensions in space but include all
three velocity dimensions. In contrast to PIC--codes, the distribution
function is integrated in time directly on a numerical grid in phase space.
The integration scheme is based on a flux conservative and positive scheme
\cite{FIL01} which obeys the maximum principle and suffers from relatively
little numerical diffusion. The main idea of the scheme is to calculate the
fluxes into and out of a grid cell by interpolating the primitive of the
distribution function. The one dimensional scheme has been generalized to
two spatial and three velocity dimensions using the backsubstitution method
described in Ref.\ \onlinecite{SCHM06b}. The backsubstitution method has been shown to
be slightly superior and, more importantly, much faster than the
straightforward time--splitting scheme.

The Maxwell equations in the Darwin approximation can be recast into a form
which does not include any time derivatives (see, for example, Refs.\
\onlinecite{BIR85,SCHM06a}). This makes it possible to express the
electromagnetic fields $\mathbf{E}(t_0)$, $\mathbf{B}(t_0)$ at any time
$t_0$ by the density $\rho(t_0)$ and the current density $\mathbf{j}(t_0)$
at the time $t_0$. No time integration of the fields is necessary and the
conditions $\nabla\cdot\mathbf{B}=0$ and $\nabla\cdot\mathbf{E}=\rho$ are
always met by construction.

\section{GEM Reconnection Setup\label{SecSetup}}

The reconnection setup is identical to the parameters of the GEM magnetic
reconnection challenge. \cite{Birn:2001a} The initial conditions are based on
the Harris sheet equilibrium \cite{HARS62} in the $x$,$y$--plane
\begin{equation}
\mathbf{B}(y) = B_0 \tanh\left( \frac{y}{\lambda} \right)\mathbf{\hat{x}}.
\end{equation}
The particles have a shifted Maxwellian distribution
\begin{equation}
f_{0 i,e}(y,\mathbf{v}) = n_0(y)
  \exp\left[ \frac{m_{i,e}}{2T_{i,e}}
    \left( v_x^2 + (v_y - V_{0 i,e})^2 + v_z^2 \right) 
  \right].
\end{equation}
Here $T_{i,e}$ are the constant electron and ion temperatures and
$V_{0 i,e}$ are the constant electron and ion drift velocities. The
density distribution is then given by
\begin{equation}
n_0(y) = n_0 \text{ sech}^2\left( \frac{y}{\lambda} \right).
\end{equation}
The Harris equilibrium demands that
\begin{alignat}{1}
\frac{B_0^2}{2\mu_0} &= n_0 (T_i + T_e)\\
\lambda    &= \frac{2}{eB_0}\frac{T_i + T_e}{V_{0i} - V_{0e}}\\
\text{and} \qquad \frac{V_{0e}}{V_{0i}} &= -\frac{T_e}{T_i}.
\end{alignat}
In addition a uniform background density $n_b = 0.2n_0$ with the same
temperature $T_{i,e}$ but without a directed velocity component is
included.

The total GEM system size is $L_x=25.6\lambda_i$ in $x$--direction and
$L_y=12.8\lambda_i$ in $y$--direction, where $\lambda_i$ is the ion
inertial length $\lambda_i = c/\omega_{pi}$ with the ion plasma frequency
is defined using the given by $\omega_{pi} = (ne^2 / \varepsilon_0
m_i)^{1/2}$. Because of the symmetry constraints we simulate only one
quarter of the total system size: $0\le x \le L_x/2$ and $0\le y \le
L_y/2$. The sheet half thickness is chosen to be $\lambda=0.5\lambda_i$.
The temperature ratio is $T_e/T_i = 0.2$ and a reduced mass ratio of
$m_i/m_e = 25$ is used. This reduced mass ratio is consistent with the GEM
setup and was chosen as a compromise between a good separation of scales
(both in space and time) and the computational effort. The numerical
expense was especially high for the kinetic codes.
\cite{Shay:01,Pritchett:2001b,Hesse:2001} The reduced mass ratio was also
used in the hybrid codes \cite{Shay:01,Kuznetsova:2001} for the sake of
comparison. A higher mass ratio (i.e. a smaller electron mass) results in a
smaller electron skin depth which has to be resolved by the grid. At the
same time the numerical time step has to be reduced to resolve the electron
Larmor frequency. Both effects together imply that, in a spatially
two--dimensional simulation, the computational cost increases with the
square of the mass ratio for all kinetic simulations independent whether
thay use PIC or Vlasov--methods. Simulations with substantially higher mass
ratios using the generally more demanding Vlasov approach, therefore,
currently seem out of reach. The simulation is performed on  $256\times128$
grid points in space for the quarter simulation box. This corresponds to a
resolution of $512\times256$. This implies a grid spacing of $\Delta x =
\Delta y = 0.05 \lambda_i$. The resolution in the velocity space was chosen
to be $20\times20\times40$ grid points. In $v_z$ direction the grid was
extended to account for the electron acceleration in the diffusion region.
The simulation was performed on a 32 processor Opteron cluster and took
approximately 150 hours to complete.

An initial perturbation 
\begin{equation}
\psi(x,y) = \psi_0\cos(2\pi x/L_x)\cos(\pi y/L_y)
\end{equation}
is added to the magnetic vector potential component $A_z$. The magnitude of
the perturbation is chosen to be $\psi_0 = 0.1 B_0/\lambda_i$. The rather
high values of the initial perturbation generates a single large magnetic
island. The initial linear growth of the tearing mode is bypassed and the
system is placed directly in the nonlinear regime. The reason for this
relatively strong perturbation is that the initial growth of the
instabilities depends strongly on the electron model. In contrast to this,
the GEM challenge demonstrated convincingly, that the later nonlinear stage
is not sensitive to the details of the underlying model.

\section{Simulation Results\label{SecResults}}

\begin{figure}[t]
\begin{center}
\includegraphics[width=8.3cm]{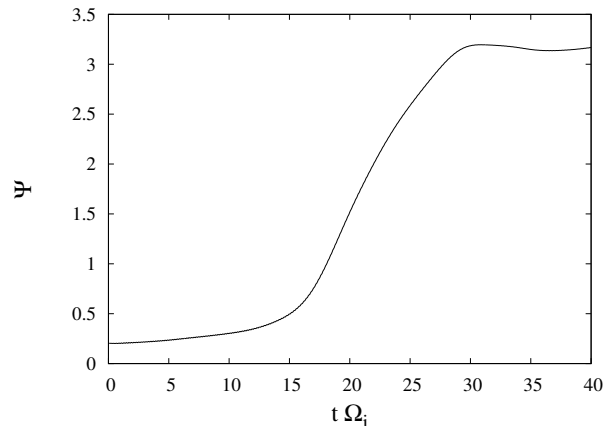}
\end{center}
\caption{Time evolution of the reconnected magnetic flux $\Psi$ throughout
the simulation run. \label{FigRecFlux}}
\end{figure}

As a measure of the reconnected magnetic field we use the difference
$\Delta \Psi$ of the magnetic vector potential component $A_z$ between the
X--point and the O--point. Figure \ref{FigRecFlux} shows the reconnected
magnetic flux as a function of time. The evolution of $\Delta \Psi$ agrees
very well with the results of other simulations of the GEM challenge.
\cite{Birn:2001a} After an initial small increase, the flux starts to rapidly
increase at $t \Omega_i \approx 15$. A value of $\Delta \Psi =
B_0/\lambda_i$ is reached at $t \Omega_i \approx 17.7$. This is slightly
later to what is observed in the PIC simulations. Pritchett
\cite{Pritchett:2001b} reports a time $t \Omega_i \approx 15.7$ where the
same flux level is reached. While the level of saturation is comparable to
the other GEM results, it is again reached slightly later than in
Ref.\ \onlinecite{Pritchett:2001b} at time $t \Omega_i \approx 30$.

\begin{figure*}
\onecolumngrid
\begin{center}
\includegraphics[width=11.5cm]{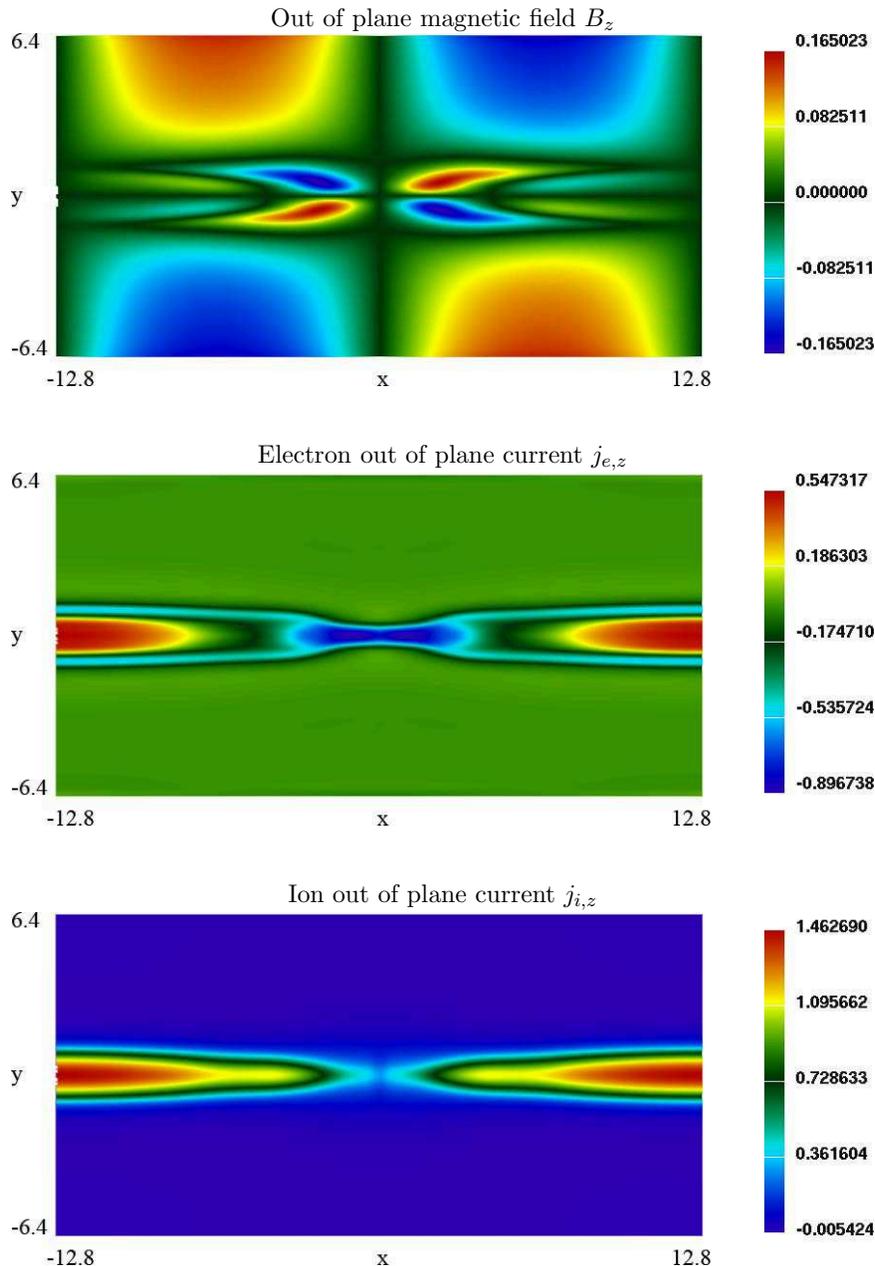}
\end{center}
\caption{The out of plane magnetic field $B_z$ (upper panel), the electron
out of plane current $j_{e,z}$ (middle panel) and the ion out of plane 
current $j_{i,z}$ (lower panel) at time $\Omega_i t =
17.7$\label{FigOutPlane}}
\twocolumngrid
\end{figure*}

The upper panel of Figure \ref{FigOutPlane} shows the out of plane magnetic
field $B_z$ at $\Omega_i t = 17.7$. One can clearly identify the
quadrupolar structure generated by the Hall currents. \cite{SONN79,TER83}
The peak value of the magnetic field $|\mathbf{B_z}|$ in the island
structure is $\approx 0.17 B_0$ and is located at $|x| \approx
1.7\lambda_i$. This is a somewhat lower magnetic field than in Ref.\
\onlinecite{Pritchett:2001b} and is also located slightly closer to the
X--line. In the course of the simulation both the size of the magnetic
island and the magnitude of the peak magnetic field increase. For this
reason, the differences to the PIC simulations can be attributed to the
small differences in the temporal behavior. This also explains the magnetic
field towards the top and bottom boundaries of the simulation box. These
fields, which can be seen in Figure \ref{FigOutPlane}, have essentially
disappeared at $\Omega_i t = 20$, approximately the same time that the peak
magnetic reconnection rate is seen.

The middle and lower panel of Figure \ref{FigOutPlane} show the electron and
ion out of plane current densities $j_{e,z}$ and $j_{i,z}$ at $\Omega_i t =
17.7$. While the ion current density almost exactly follows the ion number
density $n_i$, the electron current density is strongly enhanced near the
X--line where the number densities are depleted. The thickness of the
electron current layer is smaller than the ion skin depth but larger than
the electron skin depth. The size is determined by the meandering motion of
the electrons around the neutral line.
In addition to this current sheet, we
can observe thin current layers emanating from the X--line  which run along
the separatrix. These structures have not been reported in previous PIC
simulations but can be seen in Hall--MHD simulations. \cite{Shay:01} We
point out again that the mass ratio in the GEM simulations was fixed at
$m_i/m_e=25$ for both PIC \cite{Shay:01,Pritchett:2001b,Hesse:2001} and
hybrid \cite{Shay:01,Kuznetsova:2001} simulations. The difference between
the Vlasov code and the PIC code, therefore, can only be explained by the
differences in the numerical approach.

\begin{figure}[t]
\begin{center}
\includegraphics[width=8.3cm]{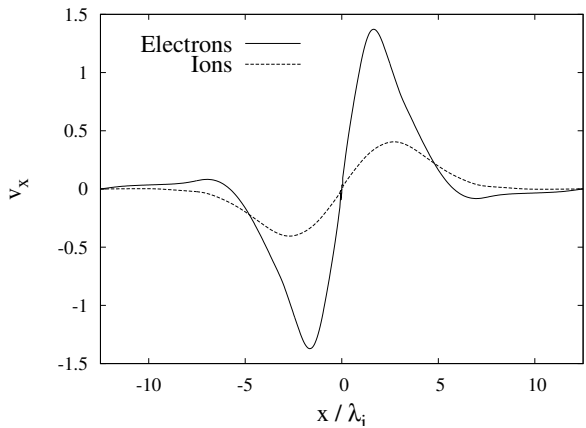}
\end{center}
\caption{Velocity profile at $\Omega_i t = 18.1$ as a function of $x$ at
the location of the current sheet $z=0$ for electrons and ions.
\label{FigEjectVel}}
\end{figure}

The electron and ion bulk velocity profiles $v_x(x)$ along the direction of
the current sheet are  shown in Figure \ref{FigEjectVel}. One can see that
the electrons are ejected away from the X--line at super--Alfv\'enic speeds.
These velocity profiles are almost identical to those reported in
Ref.\ \onlinecite{Pritchett:2001b}. In contrast to those results, no averaging
over a finite time period had to be carried out here because the Vlasov
simulations do not suffer from artificial numerical noise.

\subsection{Ohm's Law}

Within the GEM reconnection challenge it has become clear that the
Hall--MHD model is a minimal model to understand collisionless
reconnection. \cite{Shay:01} In Hall--MHD Ohm's law has the form
\begin{displaymath}
\frac{m}{ne^2}\frac{dj}{dt} = 
\mathbf{E} + \mathbf{v_i}\times\mathbf{B} 
  - \frac{1}{ne}\mathbf{j}\times\mathbf{B} 
  + \frac{1}{ne}\nabla\cdot\underline{\mathbf{P}}_e,
  \label{EqGenOhmsLaw}
\end{displaymath}
where the resistivity has been neglected. This is the exact electron momentum
equation which can be derived from kinetic theory of a collisionless plasma
without any approximations. At large scale lengths only the MHD terms play a
role, while the Hall term and the electron pressure gradient can be
neglected. To investigate the regions in which the terms of the generalized
Ohm's law become important, we calculated the different contributions in
the whole reconnection region. 

\begin{figure*}[t]
\onecolumngrid
\begin{center}
\includegraphics[width=11.5cm]{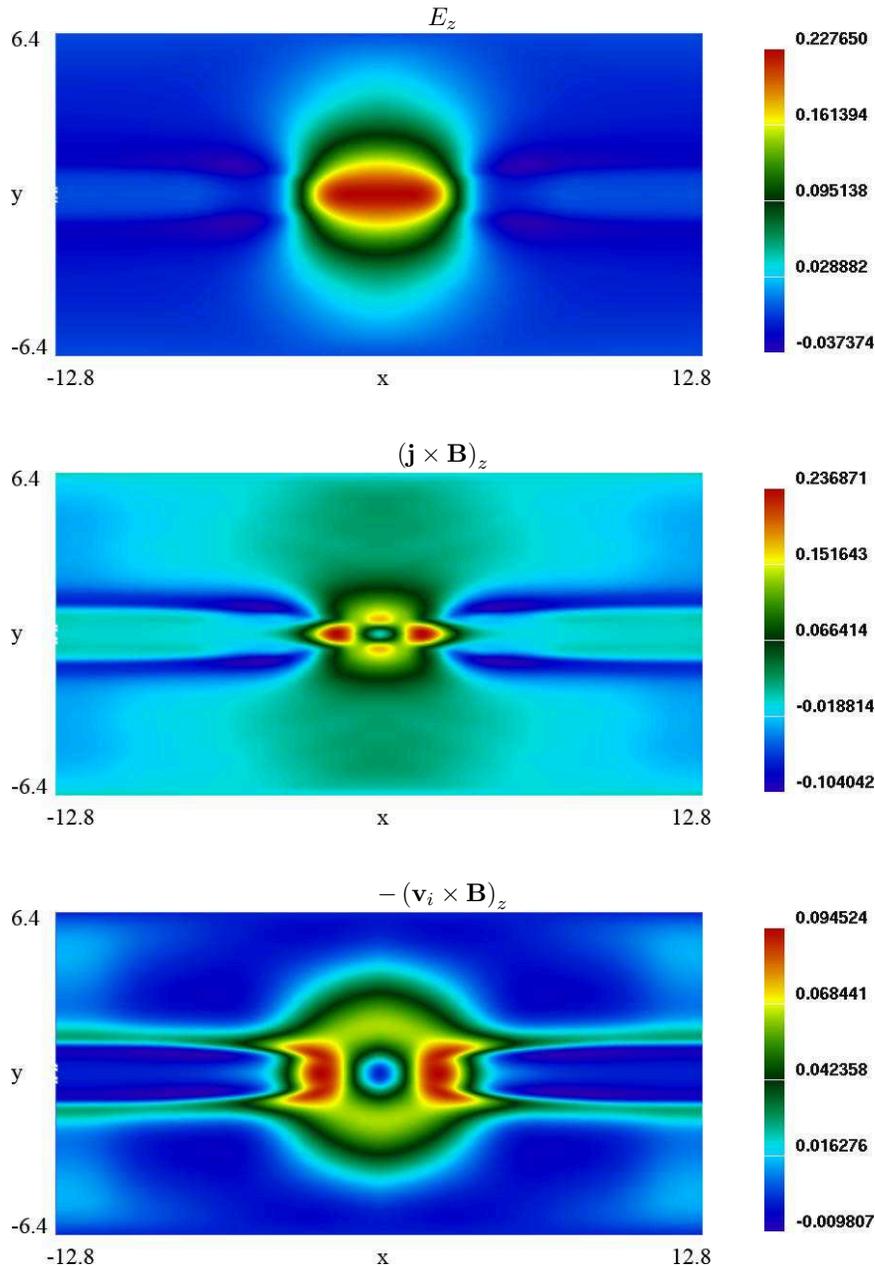}
\end{center}
\caption{$z$--component of the inductive electric field $E_z$
(upper panel), Hall Term $\mathbf{j}\times\mathbf{B}$ (middle panel) and the
negative $z$--component of $\mathbf{v}_i\times\mathbf{B}$ (lower panel)
at time $\Omega_i t = 17.7$ \label{FigOhms01}}
\twocolumngrid
\end{figure*}

The top panel of Figure \ref{FigOhms01} shows the inductive electric field
$E_z$. This inductive field is necessary for reconnection to take place. 
The region of enhanced $E_z$ is situated in a relatively large region
around the X--line. The peak electric field is located in an elongated area
extending about two ion inertial lengths in the $y$--direction and about 4
ion inertial lengths in the $x$--direction. In contrast, the region where
the field exceeds half its peak value is almost circular with a diameter of
about 5 ion inertial lengths. The middle panel of Figure \ref{FigOhms01}
shows the $z$--component of the Hall term
$\left(\mathbf{j}\times\mathbf{B}\right)_z$. In contrast to the inductive
electric field, this quantity shows a more detailed structure. Two strong
peaks are found left and right of the X--line. The peak values slightly
exceed the maximum value of $E_z$. These peaks coincide with the maxima of
the electron outflow velocity (see Figure \ref{FigEjectVel}). This shows
that the Hall term is most important in the outflow regions where the
electrons are accelerated to super Alfv\'enic velocities. In addition, two
weaker peaks are located above and below the X--line, where the electrons
are accelerated towards the X--line and the electron velocity starts to
diverge from the $\mathbf{E}\times\mathbf{B}$--velocity. Due to the
symmetry conditions, the Hall--term is exactly zero at the X--line itself.
In addition to the structure around the X--line, we also observe sheets of
negative valued Hall--term along the separatrix. We attribute this to the
current loop that generates the quadrupolar magnetic field $B_z$. Away from
the X--line, the electrons responsible for the current have to cross the
separatrix back into the upstream region in order to close the loop.
Therefore, the Hall--term will have negative values along the separatrix.
The magnitude of the Hall--term here is almost half the peak magnitude in
the X--line region. 

The bottom panel of Figure \ref{FigOhms01} shows the distribution of
$-\left( \mathbf{v}_i\times\mathbf{B} \right)_z$. This term becomes
non--zero when ions can move across the magnetic field lines in a region of
a few ion inertial lengths around the X--line. Again two peaks can be
observed in the outflow region. The peak values are, however, less than
half of the inductive electric field. A striking feature in this picture is
the almost circular ring around the X--line, where the ions become
demagnetized. The sheets of enhanced value along the separatrix are
narrower than those observed from the Hall--term. They have the same sign
as the peaks near the X--line and therefore partially cancel the
Hall--term.

\begin{figure}[t]
\begin{center}
\includegraphics[width=8.3cm]{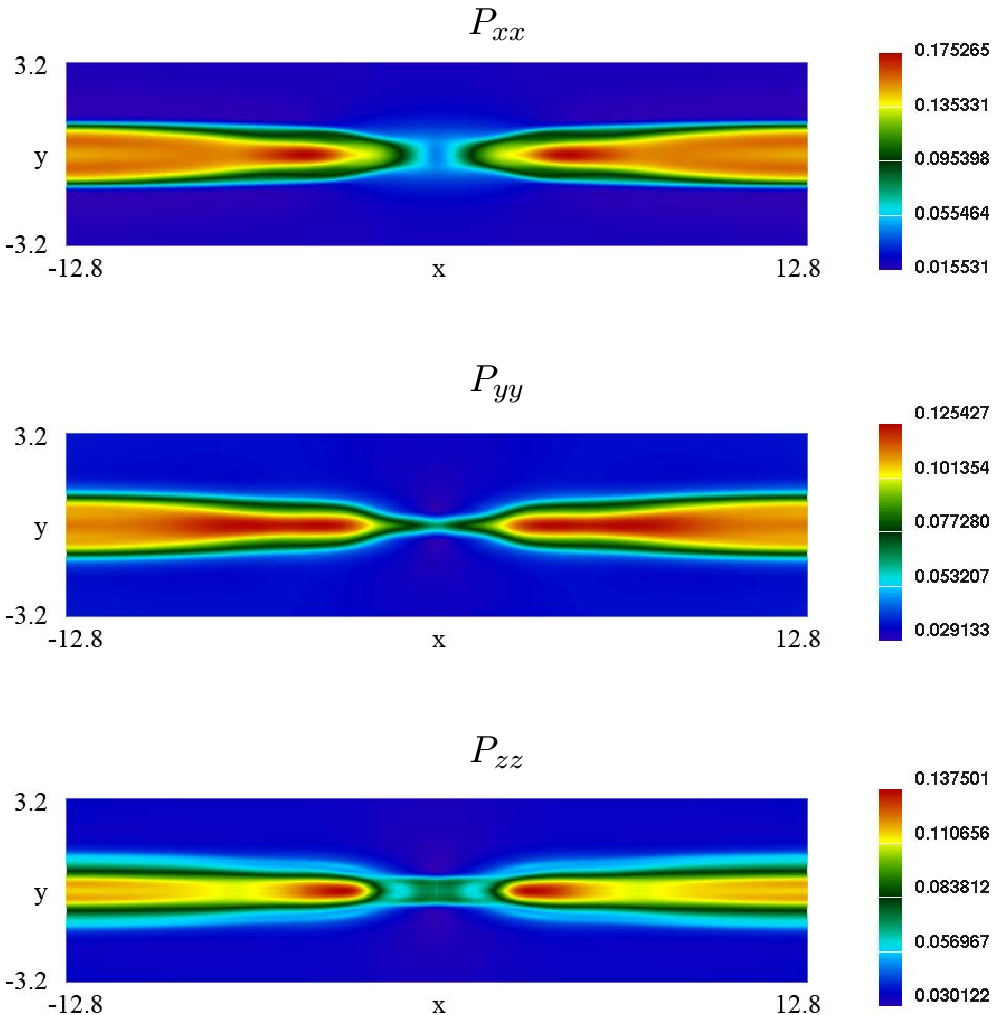}
\end{center}
\caption{The diagonal components of the pressure tensor at time $\Omega_i t
= 17.7$ \label{FigPressureDiag}}
\end{figure}

\begin{figure}[t]
\begin{center}
\includegraphics[width=8.3cm]{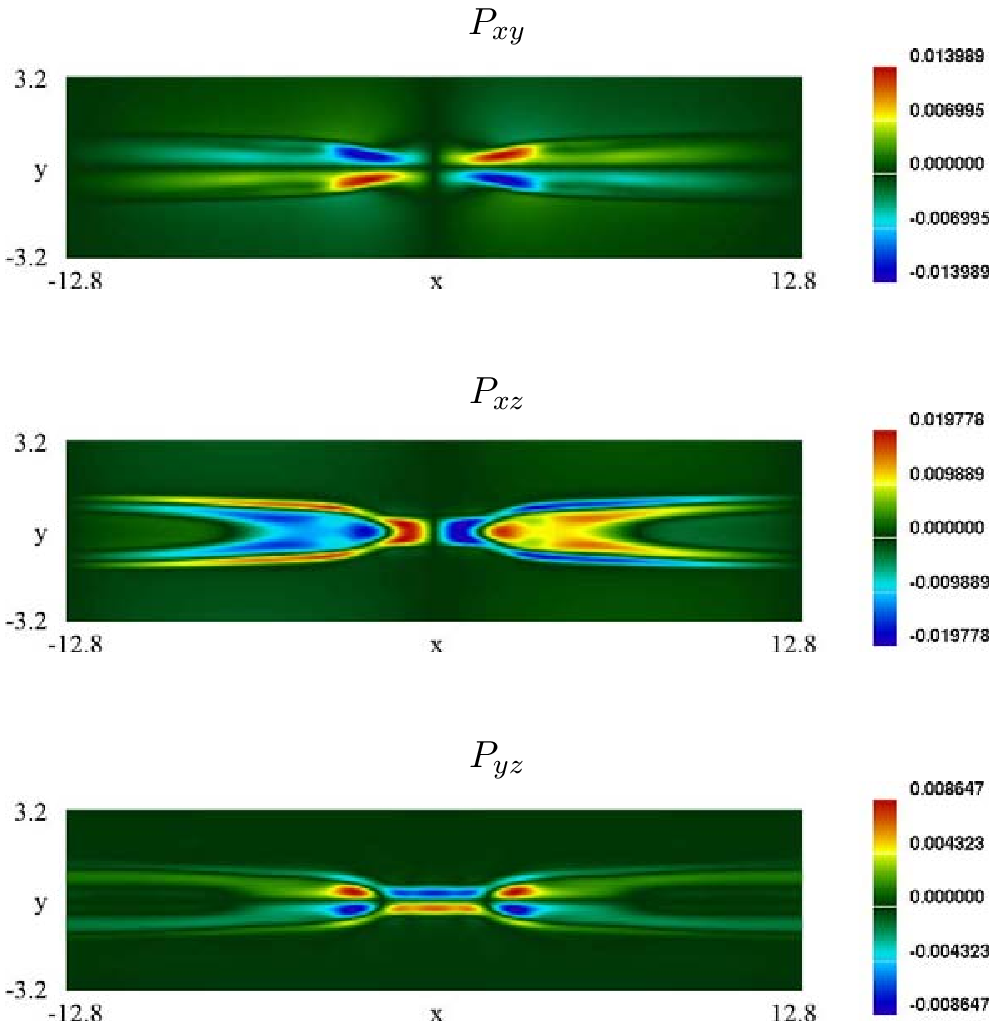}
\end{center}
\caption{The off--diagonal components of the pressure tensor at time
$\Omega_i t = 17.7$ \label{FigPressureNDiag}}
\end{figure}

Figures \ref{FigPressureDiag} and \ref{FigPressureNDiag} display the
components of the electron pressure tensor. Although only the two mixed
elements $P_{xz}$ and $P_{yz}$ play a role in the $z$--component of Ohm's
law, the other elements are shown for completeness. The upper panel of
Figure \ref{FigPressureDiag} shows the diagonal terms of the pressure
tensor.  Here we see the differences in the heating of the electrons in the
three directions. In all three cases the maximum is reached in the outflow
region, where the electron velocities are super Alfv\'enic. The heating in
the x--direction, which is mainly parallel to the magnetic field lines, is
strongest. $P_{xx}$ is increased mostly in the outflow region, with only a
slight increase in the diffusion region. In the outflow region $P_{yy}$ and
$P_{zz}$ are comparable, because these two directions are roughly
perpendicular to the magnetic field lines. Within the diffusion region
$P_{yy}$ and $P_{zz}$ are, however, different. While $P_{yy}$ is enhanced
in a narrowing X type region, $P_{zz}$ shows a bar like structure. The
reason for this structure of $P_{zz}$ may be seen in the acceleration of a
part of the electron population in the $z$--direction.  This will become
more apparent in the next section, where the electron distributions are
investigated in detail.

The off--diagonal elements of the pressure tensor are shown in Figure
\ref{FigPressureNDiag}. The magnitude of these is roughly one order of
magnitude smaller than the diagonal elements which agrees remarkably well
with the results of Kuznetsova {\it et al}. \cite{Kuznetsova:2001} The $P_{xy}$
component (top panel) shows a quadrupolar structure similar to the out out
plane magnetic field $B_z$. The $P_{xz}$ component has two extrema left and
right of the X--line at the edges of the diffusion region in agreement with
Ref.\ \onlinecite{Kuznetsova:2001}. In addition, we find two more extrema along the
$y=0$ line in the electron acceleration region and also enhanced values
around the separatrix. Finally, $P_{yz}$ shows a double bar structure in
the diffusion region and also extrema are found in the electron
acceleration region. 

\begin{figure}[t]
\begin{center}
\includegraphics[width=8.3cm]{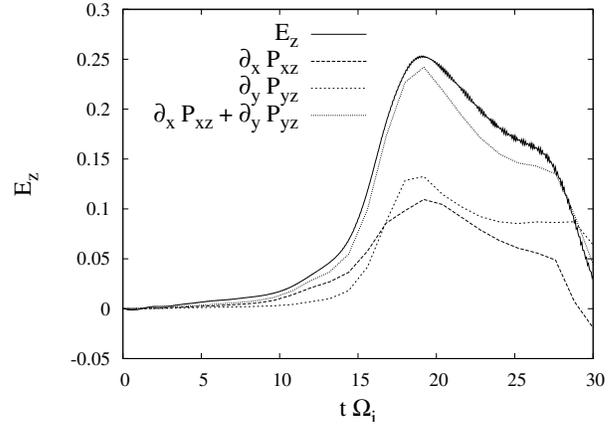}
\end{center}
\caption{Inductive electric field $E_z$ at the X--line over time together 
with the contributions $\partial P_{xz} / \partial x$ and $\partial P_{yz} / 
\partial y$ from the pressure tensor.
\label{FigEzXPoint}}
\end{figure}

The terms of the pressure tensor that contribute at the X--line $x=y=0$ are
given by $\partial P_{xz} / \partial x$ and $\partial P_{yz} / \partial y$.
In Figure \ref{FigEzXPoint} we have plotted the inductive electric field
$E_z$ at the X--line over time together with the two gradients of the
off--diagonal elements of the pressure tensor. We can clearly observe that
the two contributions $\partial P_{xz} / \partial x$ and $\partial P_{yz} /
\partial y$ are roughly equal. The sum of the two shows are remarkable
agreement with the electric field over the whole time of the simulation. 
This indicates that, at the X--line, the bulk inertia plays only a minor
role. The bulk inertia scales like $\lambda_e / L$ (see Ref.\
\onlinecite{Vasyliunas:1975}), where $\lambda_e$ is the electron inertial
length and $L$ is a typical gradient scale length. Around the zero line of
the magnetic field, the scale lengths of the electron dynamics are given
not by the electron inertial lengths, but by the larger scale of the
meandering electron motion.
\cite{Speiser:1991,Horiuchi:1997,Kuznetsova:1998} For this reason, the
contribution of the non--gyrotropic pressure exceeds the bulk electron
inertia.  For more realistic mass ratios we expect the electron bulk
inertia to be completely negligible. The pressure terms should, on the
other hand, remain important also for higher mass ratios.  Note that the
importance of the non--gyrotropic pressure has been shown only close to the
X--line. Away from the X--line, but still inside the diffusion region, we
find the Hall--term $\mathbf{j}\times\mathbf{B}$ to play a dominant role.

\subsection{Electron distribution function}

\begin{figure*}[t]
\onecolumngrid
\begin{center}
\includegraphics[width=11.5cm]{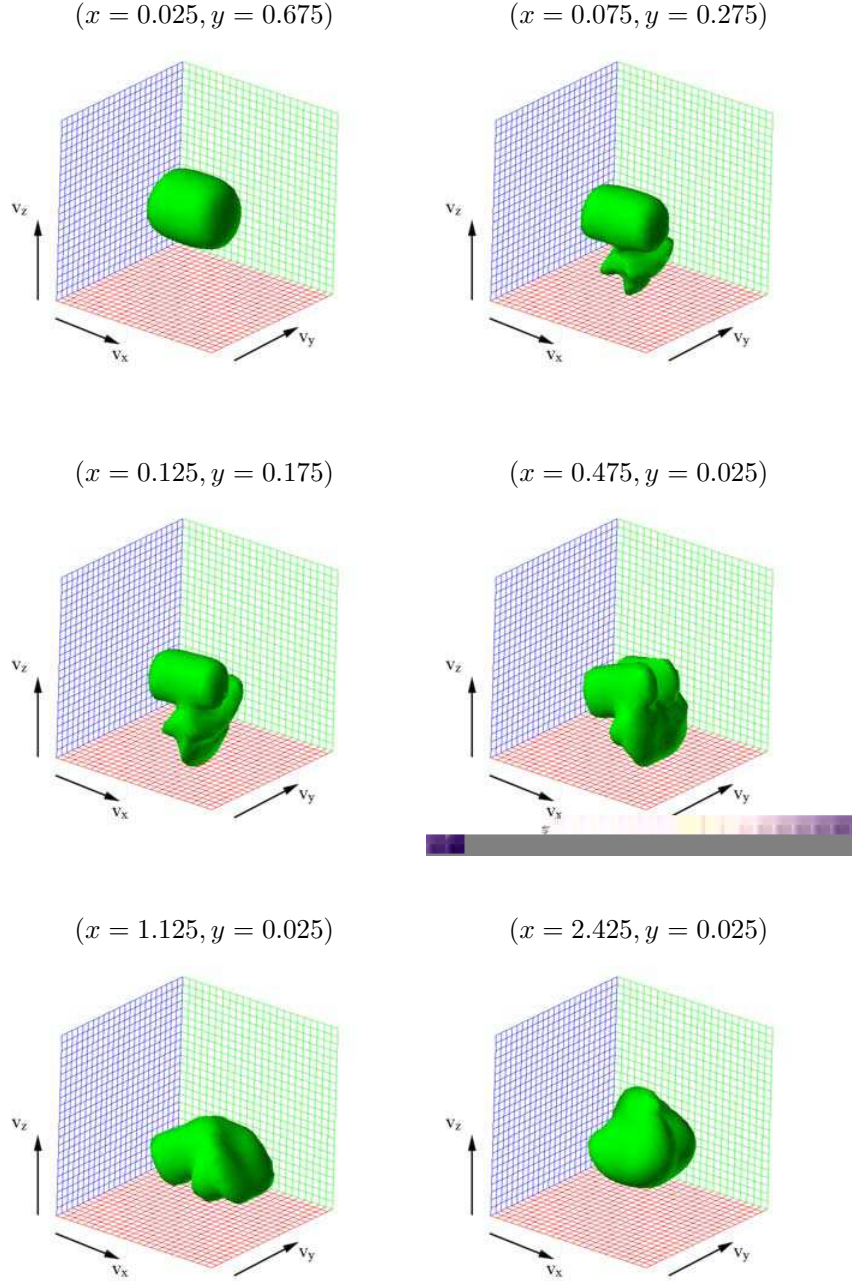}
\end{center}
\caption{Isosurface of the electron distribution function in velocity space
for different positions in the simulation box. The red plane is $v_z=$const,
the blue plane is $v_x=$const and the green plane is $v_y=$const. The
isosurface is drawn at $f_e(v_x, v_y, v_z) = f_{max}/2$ where $f_{max} =
f_{max}(x,y)$ if the maximum value of the distribution function at the
position (x,y). The velocity box ranges from $v_k =$-3.67 to 3.67 for
all three velocity components. \label{FigElDistrib}}
\twocolumngrid
\end{figure*}

To analyze the kinetic mechanism that is responsible for the generation of
the non--gyrotropic electron pressure, we have to look at the structure of
the electron distribution functions in the vicinity of the X--line. Figure
\ref{FigElDistrib} shows isosurface plots of $f_e(\mathbf{x},\mathbf{v})$
in velocity space for various fixed positions $(x,y)$ in the simulation
box. The sequence of sample points are chosen to follow a path from inflow
region, just outside the diffusion region, through the diffusion region and
to a point in the outflow region. The first panel shows the distribution
function at $(x,y)=(0.025, 0.675)$. This is close to the symmetry axis
$x=0$ and in the inflow region, just outside the diffusion region. Note
that the points could not be chosen to lie on the lines of symmetry ($x=0$
or $y=0$) since these were between the grid points in configuration space.
Here the distribution function is slightly elongated in the $x$--direction
indicating a slightly larger pressure $P_{xx} > P_{yy}, P_{zz}$ at this
point. This moderate increase in $P_{xx}$ can also be observed in Figure
\ref{FigPressureDiag}. While $P_{yy}$ and $P_{zz}$ exhibit a sharp increase
at the inflow edge of the diffusion region, $P_{xx}$ rises more gradually.
The distribution at this point is totally made up of the background
electron population. The point $(x,y)=(0.075, 0.275)$ lies just inside the
diffusion region. In addition to the original background population, here
two new electron populations suddenly appear. Both of these have a negative
$v_z$--velocity. The first population is elongated in $v_x$ and $v_y$
direction. It is slightly bent which indicates a gyro motion. We interpret
this population as being the background electrons originating from the
other side of the current sheet. These electrons move through the magnetic
zero line and show a bunched gyro motion when they enter the region on the
other side with the oppositely directed magnetic field. Finally the second
new electron population shows a sharp distribution with a strong negative
$v_z$--velocity. These are the electrons which have been accelerated in the
diffusion region.

In the left panel on the middle row of Figure \ref{FigElDistrib} the sample
point $(x,y)=(0.125, 0.175)$ has moved closer to the $y=0$ line of symmetry
but away from the $x=0$ line of symmetry. Here the  populations already
seen in the last panel become more pronounced. The gyration of the
electrons from the opposite side of the current is now more apparent. This
population has a distribution which exhibits a banana like shape, bent
around the $v_x$--axis. This indicates a bunched gyro--motion. In addition,
the distribution has been stretched to even higher negative
$v_z$--velocities. An slight asymmetry can be seen in the $v_x$--direction
due to the onsetting acceleration towards the outflow region. As one moves
away from the $x=0$ line of symmetry towards larger values of $x$, one must
distinguish between those electrons flowing into the diffusion region from
the $\pm y$--directions and those electrons that have entered the diffusion
region closer to the X--line. The latter population already has a strong
directed velocity in the outflow direction. This can be seen in the right
panel on the middle row which displays the distribution at $(x,y)=(0.475,
0.025)$. This position is almost on the $y=0$ line of symmetry. Again the
points could not be chosen to exactly lie on the line of symmetry because
of the choice of the numerical grid. In this panel, the two populations
from the $\pm y$--directions can be identified as two elongated blobs lying
parallel to each other. The fact that these two blobs are well separated
indicates that their relative velocity is larger than the thermal velocity
of the electrons.  For realistic mass ratios the theral velocity of the
electrons will increase and the structure of the distribution function will
be smeared out. Nevertheless, we expect the underlying mechanisms of
meandering motion and acceleration in the $z$--direction to remain
unchanged. The other population, which has already spent more time in the
diffusion region, has been accelerated toward the outflow region. Both, the
blobs and the accelerated population appear slightly tilted around the
$v_y$ axis. This could indicate a gyro--motion around the newly reconnected
magnetic field in the $y$--direction. In the bottom--left panel,
$(x,y)=(1.125, 0.025)$, this rotation is even more pronounced. Here the
different populations start to merge and lose their individual identity.
Finally at $(x,y)=(2.425, 0.025)$ (bottom--right panel) most of the
structure has been lost. The temperature has risen considerably and a
directed velocity in the $v_x$--direction is observed.

The strongly structured electron distribution function is responsible for
the off-diagonal terms of the electron pressure tensor. The structuring is
due to the meandering motion of the electrons in the region where the
magnetic field approaches zero and changes sign.
\cite{Horiuchi:1994,Horiuchi:1997} The distribution function is made up of
a number of distinguishable populations, which have a relative velocity,
which is higher than the thermal velocity.

\section{Summary and Conclusions\label{SecSummary}}

Two and a half dimensional Vlasov simulations were carried out on the GEM
magnetic reconnection setup. Vlasov codes have the advantage, with respect
to PIC codes, that they do not suffer from numerical noise and that the
distribution function can be analyzed with high accuracy. This advantage is
gained at the cost of a substantially higher computational effort. We could
reproduce the results of other kinetic simulations carried out of the GEM
setup \cite{Pritchett:2001b} but were also able to calculate the terms in
Ohm's law, especially the contributions from the electron pressure tensor.
This shows that, although computationally more expensive, Vlasov--codes are
a valuable tool for investigating collisionless reconnection. The large
scale structure of the magnetic field and the electron and ion current
densities agreed well with particle simulations. \cite{Pritchett:2001b} In
addition we were able to identify some small scale structures of the
electron current density which showed enhanced values along the separatrix.
While these structures were probably smeared out in PIC simulations due to
the numerical noise, they resemble more the results seen in hybrid models.
\cite{Shay:01,Kuznetsova:2001}

The analysis of the contributions to the inductive electric field in Ohm's
law show that, due to the evolution of the reconnection on ion timescales
rather than electron timescales, the bulk inertia of the electron plays a
minor role. The effect of the bulk inertia will be decreased even more, if
more realistic mass ratios are used. We could show that the Hall--term
dominates at the inflow and the outflow edges of the diffusion region.
Symmetry constraints, however, cause the Hall term and the
$\mathbf{v}_i\times B$--term to vanish at the X--line. Here we could confirm
the importance of the non--gyrotropic pressure, which was previously
investigated by Kuznetsova {\it et al}. \cite{Kuznetsova:1998,Kuznetsova:2001}
The evaluation of the gradients of the off--diagonal pressure components
throughout the whole time showed, that both $\partial P_{xz} / \partial x$
and $\partial P_{yz} / \partial y$ contribute roughly the same towards the
electric field. The sum of the two contributions could explain almost the
complete inductive electric field at the X--line.

The kinetic mechanisms responsible for the non--gyrotropic electron
pressure were uncovered by investigating the electron distribution
function. The meandering motion of the electrons in the region of the zero
magnetic field is believed to be associated with the non--gyrotropic
pressure. \cite{Speiser:1991,Horiuchi:1994,Horiuchi:1997} The complex
structure of the electron distribution function showed that the meandering
motion is responsible. For the GEM parameters it is, however, not caused by
the thermal electron motion but by the directed velocity gained in the
inflow. This may, of course, be the result of the relatively high electron
mass. In simulations with more realistic mass ratios we expect the inflow
velocity to be considerably smaller than the thermal velocity of the
electrons.

\section*{Acknowledgements}

We acknowledge the enlightening discussions with J. Dreher. This work was
supported by the SFB 591 of the Deutsche Forschungsgesellschaft. Access to
the JUMP multiprocessor computer at Forschungszentrum J\"ulich was made
available through project HBO20. Part of the computations were performed on
an Linux-Opteron cluster supported by HBFG-108-291.

\bibliographystyle{aip}

\end{document}